\documentclass[12pt]{article}
\usepackage{graphics}
\begin{document}
\title{Magnetic moments of octet baryons and sea antiquark polarizations}
\author{ Jan
Bartelski\\ Institute of Theoretical Physics,\\ Faculty of
Physics, Warsaw University,\\ Ho$\dot{z}$a 69, 00-681 Warsaw,
Poland. \\ \\ \and Stanis\l aw Tatur
\\ Nicolaus Copernicus Astronomical Center,\\ Polish Academy of
Sciences,\\ Bartycka 18, 00-716 Warsaw, Poland. \\ }
\date{}
\maketitle
\begin{abstract}
\noindent Using generalized Sehgal equations for magnetic moments
of baryon octet and taking into account $\Sigma^{0}-\Lambda $
mixing and two particle corrections to independent quark
contributions we obtain very good fit using experimental values
for errors of such moments. We present sum rules for quark
magnetic moments ratios and for integrated spin densities ratios.
Due to the $SU(3)$ structure of our equations the results for
magnetic moments of quarks and their densities depend on two
additional parameters. Using information from deep inelastic
scattering and baryon $\beta$-decays we discuss the dependence of
antiquark polarizations on introduced parameters. For some
plausible values of these parameters we show that these
polarizations are small if we neglect angular momenta of quarks.
Our very good fit to magnetic moments of baryon octet can still be
improved by using specific model for angular momentum of quarks.
\end{abstract}

\newpage

There have been several attempts
\cite{jb,jb1,karl,ger1,chli,karlip} to connect precise information
from octet baryon magnetic moments \cite{pdg} with nucleon spin
structure obtained from the analysis of polarized deep inelastic
scattering (DIS) experiments (see e.g.\cite{filipp}) and octet
baryon $\beta$-decays. There is a striking similarity  in
description of magnetic moments and axial vector couplings in
$SU(3)$ symmetrical model. In this paper we will follow
phenomenological approach of Ref.\cite{jb}.

 We modify Sehgal
equations \cite{sehg} ( justification for them was presented in
\cite{karl}) for magnetic moments of baryons given in terms of
linear independent products for $u$, $d$ and $s$ quarks of quark
magnetic moments and corresponding quark densities by taking into
account $\Sigma^{0}-\Lambda $ mixing. We also have an additional
term (we believe connected with two quark interactions) giving
contribution to nucleon magnetic moments as well as to $ \Sigma^0
\Lambda$ transition moment. The inclusion of such term enables to
satisfy Coleman-Glashow type sum rule for magnetic moments. With
these phenomenological modifications we get very good fit to octet
baryon magnetic moments using experimental values for errors.

 Unfortunately with 4 parameters used in the fit we can not
determine all 6 quantities namely magnetic moments of quarks and
quark densities. We get two relations for the ratios of magnetic
moments of quarks and for ratios of quark densities. One stands
for the ratios of magnetic moments of quarks which are not
directly measurable. Having magnetic quark densities it would be
possible, by comparing this densities with spin densities
(calculated from $\beta$-decays and DIS) to get antiquark sea
polarizations (provided that the orbital momentum contributions
are negligible). Because of the structure of our model we can only
get these quantities as a functions of two introduced by us
parameters $\epsilon$ and $g$. These new parameters are connected
with the deviation of the ratio $\mu_u/\mu_d$ from -2 and the
difference of $u$ and $d$ quark densities. We get the new sum rule
connecting antiquark sea contributions which do not depend on our
parameters. We show how sea antiquark polarizations depend on
these parameters. Assuming plausible values for these parameters
suggested by $SU(2)$ isospin symmetry and by not very conclusive
experimental data we calculate antiquark sea contributions. It is
possible still to improve our fit to magnetic moments by adding to
modified Sehgal equations term connected with orbital angular
momentum proportional to the charge of the baryon. This correction
makes fit much better and only slightly modifies obtained results.
Our phenomenological modifications of Sehgal equations are
inspired by chiral quark model \cite{fr1,fr2,ger2} with domination
of pionic exchanges.

Let start with the formulas for magnetic moments of SU(3) octet baryons in
terms of quark moments \cite{jb,ger1,sehg,carlson}:
\begin{equation}
\mu(B)=\sum_{q}\mu_{q} \,  \delta q(B) + ... \, ,
\end{equation}
\noindent where the dots represent possible collective
contributions, which will be specified later, $\mu_{q}$ is a
dipole magnetic moment of a quark $q$, whereas $\delta q(B)$
stands for integrated angular momentum density (later called
magnetic density) of $q$ flavored quark in octet baryon $B$
(beside quark spin contribution the orbital angular momentum
contribution is also possible).  The $SU(3)$ flavor symmetry
enables us to write such densities as functions of the ones in a
proton, e.g.:
\begin{eqnarray}
\delta u(n) &=& \delta d(p) \, , \nonumber \\
 \delta d(\Sigma^{+}) &=&\delta s(p) \, ,\\
 \delta s(\Lambda) &=& 2 \delta u(p)/3- \delta d(p)/3+2 \delta s(p)/3 \, , \nonumber \\
etc. \nonumber
\end{eqnarray}

In the following equations we shall use short-hand notation for
such densities in a proton $\delta q(p) \equiv \delta q$ (q=u,d,s):
\begin{eqnarray}
\mu(p) &= & \mu_{u} \,  \delta u +\mu_{d} \,  \delta d+\mu_{s} \,  \delta s \, , \nonumber \\
\mu(n) &=& \mu_{u} \,  \delta d +\mu_{d} \,  \delta u+\mu_{s} \,  \delta s \, , \nonumber \\
\mu(\Lambda) & =&  \frac{1}{6} (\mu_{u}+\mu_{d}) \,  (\delta u +4 \delta d+ \delta s)+ \frac{1}{3} \mu_{s} \,  (2\delta u -\delta d+ 2\delta s) \, , \nonumber \\
\mu(\Sigma^{+}) & =& \mu_{u} \,  \delta u +\mu_{d} \,  \delta s+\mu_{s} \,  \delta d \, , \nonumber \\
\mu(\Sigma^{0}) & =& \frac{1}{2} (\mu_{u}+\mu_{d}) \,  (\delta u + \delta s)+\mu_{s} \,  \delta d \, ,\\
\mu(\Sigma^{-}) & =&\mu_{u} \,  \delta s +\mu_{d} \,  \delta u+\mu_{s} \,  \delta d \, , \nonumber \\
\mu(\Xi^{0}) & =& \mu_{u} \,  \delta d +\mu_{d} \,  \delta s+\mu_{s} \,  \delta u \, , \nonumber \\
\mu(\Xi^{-}) & =& \mu_{u} \,  \delta s +\mu_{d} \,  \delta d+\mu_{s} \,  \delta u \, . \nonumber  \\
\mu(\Sigma^{0} \rightarrow \Lambda) &=& -\frac{(\mu_{u}-\mu_{d}) \, ( \delta u - 2\delta d+ \delta s)}{2\sqrt{3}} \, . \nonumber
\end{eqnarray}

In the nonrelativistic quark model (NQM) with the $SU(6)$ symmetric wave function one gets for spin densities:
\begin{equation}
\delta u = \frac{4}{3} \; , \; \delta d =  -\frac{1}{3} \; , \; \delta s=0 \; .
\end{equation}

Because we have postulated the $SU(3)$ symmetry  the formulas for magnetic moments can be written
with 4 parameters only, instead of three moments $\mu_u$, $\mu_d$, $\mu_s$
and three densities $\delta u$, $\delta d$ and $\delta s$, namely:
\begin{eqnarray}
\mu(p) &= & c_{0}+2c_{8}+2c_{3} \, , \nonumber \\
\mu(n) &=& c_{0}+2c_{8}-2c_{3} \, , \nonumber \\
\mu(\Lambda) & =& c_{0}-(3r-1)c_{8} \, , \nonumber \\
\mu(\Sigma^{+}) & =& c_{0}+(3r-1)c_{8}+(1+1/r)c_{3} \, , \nonumber \\
\mu(\Sigma^{0}) & =& c_{0}+(3r-1)c_{8} \, ,\\
\mu(\Sigma^{-}) & =&c_{0}+(3r-1)c_{8}-(1+1/r)c_{3} \, , \nonumber \\
\mu(\Xi^{0}) & =& c_{0}-(3r+1)c_{8}-(1-1/r)c_{3} \, , \nonumber \\
\mu(\Xi^{-}) & =& c_{0}-(3r+1)c_{8}+(1-1/r)c_{3} \, , \nonumber  \\
\mu(\Sigma^{0} \rightarrow \Lambda) &=& -\frac{(3-1/r)c_{3}}{\sqrt{3}} \, ,\nonumber
\end{eqnarray}
where
\begin{eqnarray}
c_{0} &= & (\mu_{u}+\mu_{d}+\mu_{s})(\delta u+\delta d+\delta s)/3 \, , \nonumber \\
c_{3} &=& (\mu_{u}-\mu_{d})(\delta u-\delta d)/4 \, ,\\
c_{8} & =& (\mu_{u}+\mu_{d}-2\mu_{s})(\delta u+\delta d-2\delta s)/12 \, , \nonumber \\
r & =& \frac{\delta u-\delta d}{\delta u +\delta d-2\delta s} \, . \nonumber
\end{eqnarray}
The last parameter takes in NQM the value: $r=5/3$.

All magnetic moments in Eq.(5) have the same scalar ($c_0$) contribution, differ by a hypercharge ($c_8$)
and isovector ($c_3$) terms, whereas within isospin multiplet differ only by a sign of an isovector contribution.

As far we have the experimental data \cite{pdg} for seven magnetic moments
and for one transition moment $\mu(\Sigma^{0} \rightarrow \Lambda)$ (see Table I).

Because one has seven measured magnetic moments and only four parameters we can write in our model three sum rules.
One is the isovector (of Coleman-Glashow type):
\begin{equation}
[\mu(\Sigma^{+})-\mu(\Sigma^{-})]-[\mu(\Xi^{0})-\mu(\Xi^{-})]=\mu(p)-\mu(n) \, .
\end{equation}
Left hand-side of this sum rule gives $4.22\pm 0.03$ n.m, whereas
the right hand-side $4.71\pm 0.00$ n.m. It means that it is not
possible to satisfy this sum rule with our formulas with
independent quark contributions only as was already mentioned
before \cite{jb,karl,ger2,fr3}. We can save this sum rule adding
an isovector contribution to nucleon moments (such contribution
can arise e.g. when one considers charge pion exchange between
different quarks  \cite{jb,fr1,fr2,ger2}) and transition moment.
It was stressed by Franklin \cite{fr1,fr2} that the exchange of
charged pion between different quarks gives additional correction
to magnetic moment of baryon (such correction is not of one
particle type contribution like the terms in Eq.(3)). This
correction  connected with exchange of charged pion requires the
presence of $u$ and $d$ quarks in the baryon and gives
contribution only to proton, neutron and transition moment
$\Sigma\rightarrow \Lambda$. Hence, we add a new contribution to
formulas in Eq.(5) (dots represent the terms already written):
\begin{eqnarray}
\mu(p)&=&\ldots+V \, ,\nonumber \\
\mu(n)&=&\ldots -V \, ,\\
\mu(\Sigma^{0} \rightarrow \Lambda)&=&\ldots-\frac{1}{\sqrt{3}}V \, ,\nonumber
\end{eqnarray}
In principle for $\Sigma^{0} - \Lambda$ transition moment independent parameter could be introduced but we assume that it is the same V
 with a coefficient given by the SU(6) symmetrical wave function. Hence,  $\mu(\Sigma^{0} - \Lambda)$ can be predicted from our fit.

The isoscalar sum rule (of Gell-Mann-Okubo type) reads:
\begin{equation}
3\mu(\Lambda)+[\mu(\Sigma^{+})+\mu(\Sigma^{-})]/2
=[\mu(p)+\mu(n)]+[ \mu(\Xi^{0})+\mu(\Xi^{-})]
\end{equation}
Left hand-side of this sum rule gives $-1.19\pm 0.02$ n.m. whereas
right hand-side $-1.02\pm 0.01$ n.m. However, it has been pointed
out \cite{fr1} that $\Sigma^{0} - \Lambda$ mixing should be taken
into account if one considers octet baryon magnetic moments.
Defining :
\begin{eqnarray}
|\Lambda> &=& \hspace*{0.39cm} \cos \alpha|\Lambda_{SU(3)}>+\sin \alpha |\Sigma^{0}_{SU(3)}> \, ,  \nonumber \\
|\Sigma^{0}> &=& -\sin \alpha |\Lambda_{SU(3)}>+\cos \alpha |\Sigma^{0}_{SU(3)}> \, .
\end{eqnarray}
with  $ \tan \alpha $ calculated from hyperon mass differences  \cite{fr1,fr4}:
\begin{equation}
t \equiv \tan \alpha  \approx 0.014 \pm 0.004 \, ,
\end{equation}
one gets (using the experimental numbers from Table I and Eq.(11)) for
$\mu(\Lambda_{SU(3)})= c_0-(3 r-1) c_8$:
\begin{equation}
\mu(\Lambda_{SU(3)}) \approx \mu(\Lambda)- 2t
\mu(\Sigma^{0} \rightarrow \Lambda) = -0.568 \pm 0.014 \, \mbox{n.m.}\; .
\end{equation}
Inserting this number in the left hand-side of an isoscalar formula we get $-1.06\pm 0.04$ n.m. for left hand-side  which gives a good agreement
of both sides of such sum rule. It is also possible to obtain parameter of $\Sigma^{0} - \Lambda$ mixing from magnetic moments of baryons
\cite{ger1} reducing however number of degrees of freedom.

The third sum rule can be chosen as:
\begin{eqnarray}
[\mu(\Sigma^{+})+\mu(\Sigma^{-})][\mu(\Sigma^{+})-\mu(\Sigma^{-})] & - &[\mu(\Xi^{0})+\mu(\Xi^{-})][\mu(\Xi^{0})-\mu(\Xi^{-})] \nonumber \\
=[\mu(p)+\mu(n)]\{ [(\mu(\Sigma^{+})-\mu(\Sigma^{-})]&-&[\mu(\Xi^{0})-\mu(\Xi^{-})] \} \, .
\end{eqnarray}
The right hand-side gives $(1.89\pm 0.02 \, \mbox{n.m.})^2$, whereas the left $(1.93\pm 0.01\, \mbox{n.m.})^2$ with a good agreement between them.

The fact that sum rules for magnetic moments can not be satisfied with the expressions given by generalized Sehgal equations was pointed out by
several authors \cite{jb, karl,fr3} and hence, there was a problem in getting good fit to magnetic moments (artificial errors were introduced).

In our fit we use 6 measured magnetic moments of octet baryons and
$\mu(\Lambda_{SU(3)})$ from Eq.(12) with the experimental errors
getting very good result for $\chi^{2}/d.o.f. = 1.3$. Our fitted
parameters are:
\begin{eqnarray}
c_{0} &= & 0.054 \pm 0.001\, n.m. \, , \nonumber \\
c_{3} &=& 1.046 \pm 0.005\, n.m. \, , \nonumber \\
c_{8} & =& 0.193 \pm 0.000\, n.m. \, , \\
r & =& 1.395 \pm 0.010\,  , \nonumber \\
 V &=& 0.26 \pm 0.01\, n.m. \; . \nonumber
\end{eqnarray}

The results for  magnetic moments, as well as predictions for $\mu(\Sigma^{0})$ and
$\mu(\Sigma^{0} \rightarrow \Lambda)$:
\begin{eqnarray}
\mu(\Sigma^{0}) &= & c_0+ (3r-1)c_8+2t\frac{(3r-1)c_3+r V}{\sqrt{3}r}+O(t^2) \, , \nonumber \\
\mu(\Sigma^{0} \rightarrow \Lambda)& = &-\frac{(3r-1)}{\sqrt{3}r}c_3 -\frac{1}{\sqrt{3}}V+2t(3r-1)c_8+O(t^2) \, ,
\end{eqnarray}
\noindent are presented in Table \ref{tab1} (model A).

\vspace{0.2cm}
\begin{table}[h]
\caption{\label{tab1}Magnetic moments of octet baryons.}
\begin{tabular}{c r r r} \hline \hline
magnetic & experiment   \hspace*{0.4cm}& model A \hspace*{0.5cm}&
model B \hspace*{0.45cm}\\
 moment &(n.m) \hspace*{0.9cm}&(n.m) \hspace*{0.7cm} &(n.m)  \hspace*{0.7cm}\\ \hline
$\mu(p$)&$2.792847351(28)$& $2.793 \pm 0.000$& $2.793 \pm 0.000$\\
$\mu(n)$ & $-1.91304273(45)$ &$-1.913 \pm 0.000$ &$-1.913 \pm
0.000$\\ $\mu(\Lambda)$ &$ -0.613\pm 0.004$ & $ -0.606 \pm 0.015$
& $ -0.610 \pm 0.014$\\ $\mu(\Sigma^{+})$ &$2.458 \pm 0.010$ &
$2.465 \pm 0.014$ & $2.458 \pm 0.010$\\ $\mu(\Sigma^{-})$ &
$-1.160 \pm 0.025$ & $-1.128 \pm 0.013$& $-1.160 \pm 0.025$ \\
$\mu(\Xi^{0})$ & $ -1.250 \pm 0.014$ & $1.243 \pm 0.015$ & $1.252
\pm 0.012$ \\ $\mu(\Xi^{-})$ & $ -0.6507 \pm 0.0025$ &$ -0.651 \pm
0.002$ &$ -0.651 \pm 0.002$ \\ \hline $\mu(\Sigma^{0})$ & $
?\hspace*{1.1cm}$ & $0.714 \pm 0.014$ & $0.694 \pm 0.019$ \\
$\mu(\Sigma \rightarrow \Lambda)$ & $ -1.61 \pm 0.08$
\hspace*{0.08cm}&$ -1.51 \pm 0.01$ \hspace*{0.08cm}&$ -1.51 \pm
0.02$ \hspace*{0.08cm}\\ \hline \hline
\end{tabular}
\end{table}

\vspace{0.3cm}

If we do not add the isovector contribution to the transition
moment one gets (in model A): $\mu(\Sigma \rightarrow
\Lambda)=-1.36 \pm 0.01$. Hence, the inclusion of this correction
(see Eq.(8)) improves an agreement with the experimental number.

If we use in our calculations NQM densities from Eq.(4)  we get much worse fit with
$\chi^2/d.o.f. \approx 338$ and the same is true ($\chi^2/d.o.f. \approx 212$) if we neglect  isovector contribution (i.e., if we put $V=0$).
If we do not take into account  $\Sigma^{0} - \Lambda$ mixing the resulting fit gives $\chi^2/d.of. \approx 27$.
In all fits the relaxation of the condition $r=5/3$ gives (no matter if we have $V=0$ or $V \neq 0$) for this parameter the values in a range
1.38-1.48 far from the value gotten from octet baryon $\beta$ decays: $2.13 \pm 0.10$.

In formulas (8) taking into account Eqs.(5) and (6) we have three
magnetic moments of quarks and three magnetic quark densities and
only four fitted parameters. We can not calculate magnetic moments
of quarks and magnetic quark densities without additional
assumptions. As the result of our model for magnetic moments we
can only get relations for the ratios of magnetic moments of
quarks and magnetic quark densities:

\begin{equation}
\frac{\mu_s}{\mu_d}=\frac{c_{3}+3rc_{8}}{2 c_{3}}+
\frac{c_{3}-3rc_{8}}{2c_{3}}\frac{\mu_u}{\mu_d}
\end{equation}

Assuming that $\mu_u/\mu_d=-2$ (one gets this result e.g. taking
Dirac magnetic moments
 for light quarks with equal masses) we get $\mu_s/\mu_d=0.66 \pm 0.01$, the
value which is consistent with the NQM result.

For the ratios of magnetic quark densities we get:
\begin{equation}
\frac{\delta u}{\delta d}=-\frac{r+1}{r-1}+\frac{2
r}{r-1}\frac{\delta s}{\delta d}\, .
\end{equation}
For $\delta s=0$ (NQM assumption) we get $\delta u/ \delta d =
-6.1 \pm 0.1$, the number far from the naive result which is -4
(see Eq.(4)).  In this case we get for the ratio
$\mu_u/\mu_d=-1.83 \pm 0.01$. To verify our model one has to check
whether our sum rules Eqs.(16,17)  are satisfied.

 In order to determine
magnetic moments of quarks and magnetic moment densities we
introduce two additional parameters. It is convenient to take them
as $\epsilon$ and $g$:

\begin {eqnarray}
\epsilon &=&-1-2\frac{\mu_{d}}{\mu_{u}} \; , \nonumber \\
g &=& \delta u-\delta d \; .
\end{eqnarray}

We have chosen parameter  $\epsilon$ describing deviation from
isotopic $SU(2)$ symmetry for magnetic moments (that we do not
expect to be broken very strongly) and an analog of $g_A$ for
magnetic quark densities.

Using the function:
 \begin{equation}
 f(\epsilon)=\frac{(3+\epsilon) r c_{0}}{c_{3}-3rc_{8}-\epsilon (c_{3}+rc_{8})} \, ,
 \end{equation}
 we can write formulas for our densities (which depend on $\epsilon$ and $g$) in simple form:
 \begin{eqnarray}
 \delta u &=& \frac{g}{6r}[f(\epsilon)+1+3r] \, , \nonumber \\
 \delta d &=& \frac{g}{6r}[f(\epsilon)+1-3r] \, , \\
 \delta s &=& \frac{g}{6r}[f(\epsilon)-2] \, ,\nonumber
 \end{eqnarray}
which became even simpler if we use $SU(3)$ densities (i.e., scalar, hypercharge and isovector):
\begin{eqnarray}
\delta_{0} &\equiv & \delta u+\delta d+\delta s=\frac{gf(\epsilon)}{2r} \, , \nonumber \\
\delta_{8} &\equiv & \delta u+\delta d-2\delta s=\frac{g}{r} \, , \\
\delta_{3} &\equiv & \delta u-\delta d=g \, . \nonumber
\end{eqnarray}
From  Eq.(21) we see that the function $f(\epsilon)$ has an
interpretation of the ratio 2$\delta_0/\delta_8$.

The formulas for quark magnetic moments are:
\begin{eqnarray}
 \mu_{u} &=& \frac{2r}{g}[\frac{c_{0}}{f(\epsilon)}+c_{8}+\frac{c_{3}}{r}] = \frac{8 c_3}{g(3+\epsilon)}\, , \nonumber \\
 \mu_{d} &=& \frac{2r}{g}[\frac{c_{0}}{f(\epsilon)}+c_{8}-\frac{c_{3}}{r}]=-\frac{4 (1+\epsilon) c_3 }{g(3+\epsilon)} \, , \\
 \mu_{s} &=& \frac{2r}{g}[\frac{c_{0}}{f(\epsilon)}-2c_{8}]=-\frac{2[9 r c_8- c_3 +\epsilon(c_3+3rc_8)]}{g(3+\epsilon)} \, . \nonumber
 \end{eqnarray}
One can see from Eqs.(20) and (22) that the differences of the
quantities $\delta q - \delta {q'}$ and $\mu_q -  \mu_{q'}$ do not
depend on $\epsilon$ whereas the ratios of these quantities:
$\delta q / \delta q'$ and $\mu_q / \mu_{q'}$ on parameter $g$.
The quantities $\mu_q \delta q$ are also scale
 independent, i.e. does not depend on parameter $g$.

 It is not obvious what value of parameter $\epsilon$ one should use.
Because mass of the $u$ quark is different from the $d$ quark
and/or that in some models mesonic and gluonic corrections change
the values of quark magnetic moments one would expect $\epsilon$
to be slightly different from zero. In principle it is possible to
get information on the $ \mu_{u}/ \mu_{d}$ ratio from other
sources, e.g. from radiative vector meson decays but unfortunately
experimental data and theoretical framework are not accurate
enough. Not knowing the precise value of $\epsilon$ we will
present the results for two values of this parameter, namely
$\epsilon=0$ ($f(\epsilon)=0.94\pm 0.04$ in such case) and as an
example $\epsilon=\epsilon_0=0.093\pm 0.006$
($f(\epsilon=\epsilon_{0})=2$ and $\delta s=0$ in this case).
However, the zeroth approximation will be the choice $\epsilon=0$
and $g=g_{A}$.

Now we will try to determine sea antiquark polarizations. Our
magnetic densities:
\begin{equation}
\delta q=\Delta q_{val}+\Delta
q_{sea}-\Delta\bar{q}+<\hat{L}_{z}^q>-<\hat{L}_{z}^{\bar q}>,
\end{equation}
can be expressed by valence ($\Delta q_{val}$), sea quark ($\Delta
q_{sea}$) and sea antiquark ($\Delta \bar{q}$) contributions
\cite{jb,chli,carlson} only (if we {\it neglect} orbital momenta
strictly speaking if we put $<\hat{L}_{z}^q>=<\hat{L}_{z}^{\bar
q}>$ for all flavors):
 \begin{equation}
 \delta q=\Delta q_{val}+\Delta q_{sea}-\Delta\bar{q} \, .
 \end{equation}
The axial spin densities, used in DIS analysis, differ by a sign
in an antiquark term:
\begin{equation}
 \Delta q=\Delta q_{val}+\Delta q_{sea}+\Delta\bar{q} \, .
 \end{equation}
 It is clear from Eqs.(24) and (25) that if we knew the axial
 quark densities and magnetic quark densities calculated from
 magnetic moments we could calculate antiquark polarizations
 $\Delta \bar{q}$.
 The quantities $ \Delta q$ are usually expressed by the scalar, hypercharge and isovector axial charges:
 \begin{eqnarray}
 \Delta u &=& \frac{1}{3}\Delta \Sigma+\frac{1}{6}a_{8}+\frac{1}{2}g_{A} \, , \nonumber \\
 \Delta d &=& \frac{1}{3}\Delta \Sigma+\frac{1}{6}a_{8}-\frac{1}{2}g_{A} \, ,\\
 \Delta s &=& \frac{1}{3}\Delta \Sigma-\frac{1}{3}a_{8} \, . \nonumber
\end{eqnarray}

In numerical calculations we use $g_{A}=1.2695 \pm 0.0029$ \cite{pdg}, whereas we take a value of  $a_{8}$ from our fit to the
experimental data \cite{pdg} on $\beta$ decays of neutron and hyperons. We use the data for following $\beta$ decays: $g_{A} \equiv g_{A}/g_{V}(n \rightarrow p)$,
$g_{A}/g_{V}(\Xi^{-} \rightarrow \Lambda)+g_{A}/g_{V}(\Lambda \rightarrow p)$ (we use the sum of these quantities in
order to get rid of eventual corrections from $\Sigma^{0} - \Lambda$ mixing), $g_{A}/g_{V}(\Sigma^{-} \rightarrow n)$ and
$g_{A}/g_{V}(\Xi^{0} \rightarrow \Sigma^{+})$. From such a fit, with very good $\chi^2/d.o.f.=0.27$, we get: $a_{8}=0.597 \pm 0.029$.

From Eqs.(24) and (25) we have:
\begin{equation}
\Delta\bar{q}= \frac{1}{2}(\Delta q-\delta q) .
 \end{equation}
Our sum rule Eq.(17) can now be rewritten in terms of antiquark
polarizations.
\begin{equation}
 \Delta \bar{u}- \Delta \bar{d}-\frac{2r}{r+1} ( \Delta \bar{u}- \Delta \bar{s})=\frac{g_{A}-ra_{8}}{2(r+1)} \, .
\end{equation}
That is the only result we can get in our model for antiquark
polarizations  without additional assumptions. One can see that we
can not have in our model all $\Delta \bar{q}$ equal to zero (when
$<\hat{L}^{q}_{z}>-<\hat{L}^{\bar q}_{z}>=0$) because of the term
on the right-hand side of Eq.(28) (which numerical value is $0.09
\pm 0.01$). If we knew the antiquark polarizations
$\Delta\bar{u}$, $\Delta\bar{d}$ and $\Delta\bar{u}$ we could
check whether this sum rule is satisfied in  our model for
magnetic moments.

It seems that data on magnetic moments are much more precise so we
will work in the opposite direction and try to find out how the
information on magnetic moments of baryons could be used to
estimate antiquark polarizations $\Delta\bar{q}$. If we knew the
values of $g$ and $\epsilon$ we would determine the sea antiquark
polarizations.

There are several results, coming from fits to the experimental
data, for the value of $\Delta \Sigma$. They lie in a range
between 0.2 \cite{sig} and  0.4 or even 0.45 \cite{BT2}. To take
into account the spread of these values we assume $\Delta
\Sigma=0.3 \pm 0.1$ (an artificial error represents our
uncertainty of this quantity). Taking this value we get from
Eq.(25):
\begin{eqnarray}
 \Delta u &=& \hspace*{0.33cm}0.83 \pm 0.03 , \nonumber \\
 \Delta d &=& -0.44 \pm 0.03 ,\\
 \Delta s &=&  -0.10 \pm 0.03 .\nonumber
\end{eqnarray}
Unfortunately from inclusive deep inelastic scattering of
polarized particles we can not get splitting of axial quark
densities into quark and antiquark contributions. The sensitivity
to each individual quark flavor can be realized in semi-inclusive
deep inelastic scattering (SIDIS) in which the leading hadron is
also detected.

 Let us introduce the new parameter
$\eta$ (instead of $g$) defined as:
\begin{equation}
\eta=\frac{1}{2}(g_{A}-g) \, ,
\end{equation}
which gives the difference of $\bar{u}$ and $\bar{d}$ polarizations, i.e. $\eta = \Delta \bar{u}-\Delta \bar{d}$. This quantity
has been measured by the HERMES collaboration \cite{her} in semi-inclusive DIS.
The result  is $\Delta_{H}=\int^{0.3}_{0.023} [\Delta \bar{u}(x)-\Delta \bar{d}(x)]dx =
0.05 \pm 0.07$ (an extrapolation to the whole $x$ region, i.e. for $0\leq x\leq 1$ with vanishing  $\Delta_{H}(x)$ at $x=0$
and $x=1$ gives $\eta \approx 0.1$). The wide spectrum of theoretical models
 giving different values for $\eta$ are presented in \cite{peng}.
 A new Jefferson Lab experiment \cite{jeff} on spin flavor decomposition is planned
 and
 $\Delta u_{v}$,  $\Delta d_{v}$ and $\Delta\bar{u}-\Delta\bar{d}$
 will be extracted from the measurement of the combined
 asymmetry $A_{1N}^{\pi^{+}-\pi^{-}}$.
One can hope that the value of $\eta$ will be known with better
accuracy in the near future. As was already mentioned before we do
not know the precise value of $\mu_u /\mu_d$ (and in consequence
$\epsilon$) or $\mu_s /\mu_d$ ( Eq. (16)) and how can one measure
it. In principle this information is contained in radiative decays
of vector into pseudoscalar mesons but the data have big errors
and are not very consistent. Not knowing the precise value of
parameter $\eta$ (having only suggestion from Hermes experiment)
and not knowing the precise value of parameter $\epsilon$ (but
expecting that isotopic $SU(2)$ symmetry for $u$ and $d$ quarks
should not be strongly broken so $\epsilon$ will be not very much
different from zero) we discuss the dependence of antiquark
polarizations on these parameters near the above mentioned values.

\begin{figure}
\noindent \hspace{-0.5cm} \scalebox{0.7}
{\includegraphics{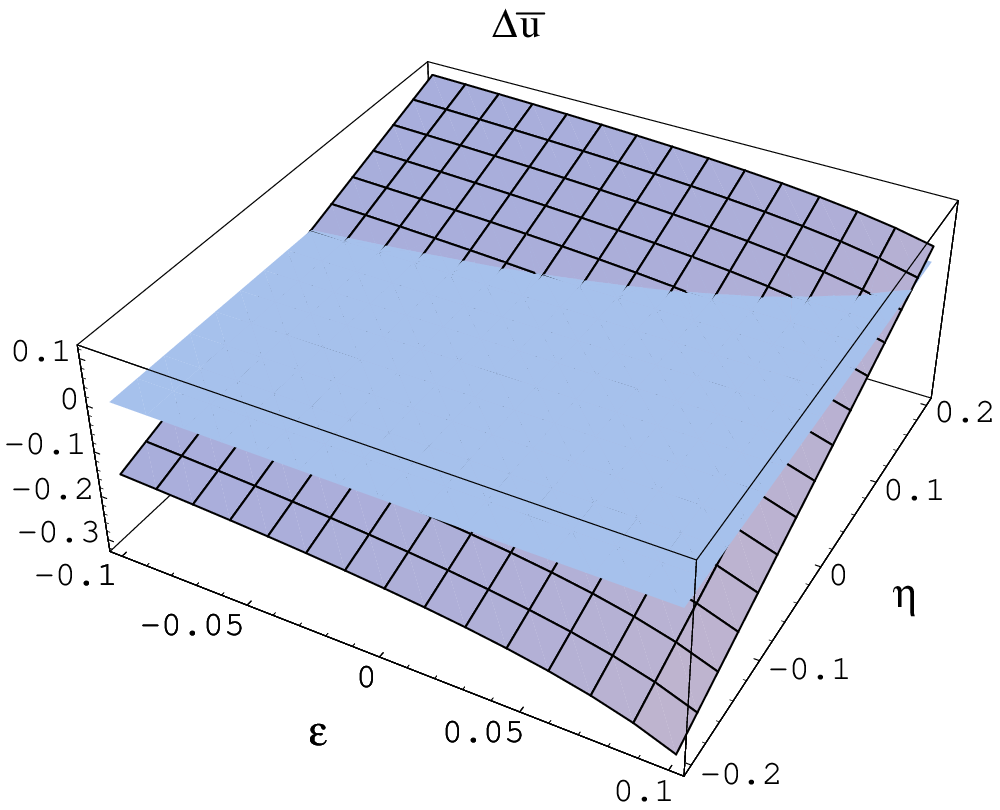}} \scalebox{0.7}{
\includegraphics{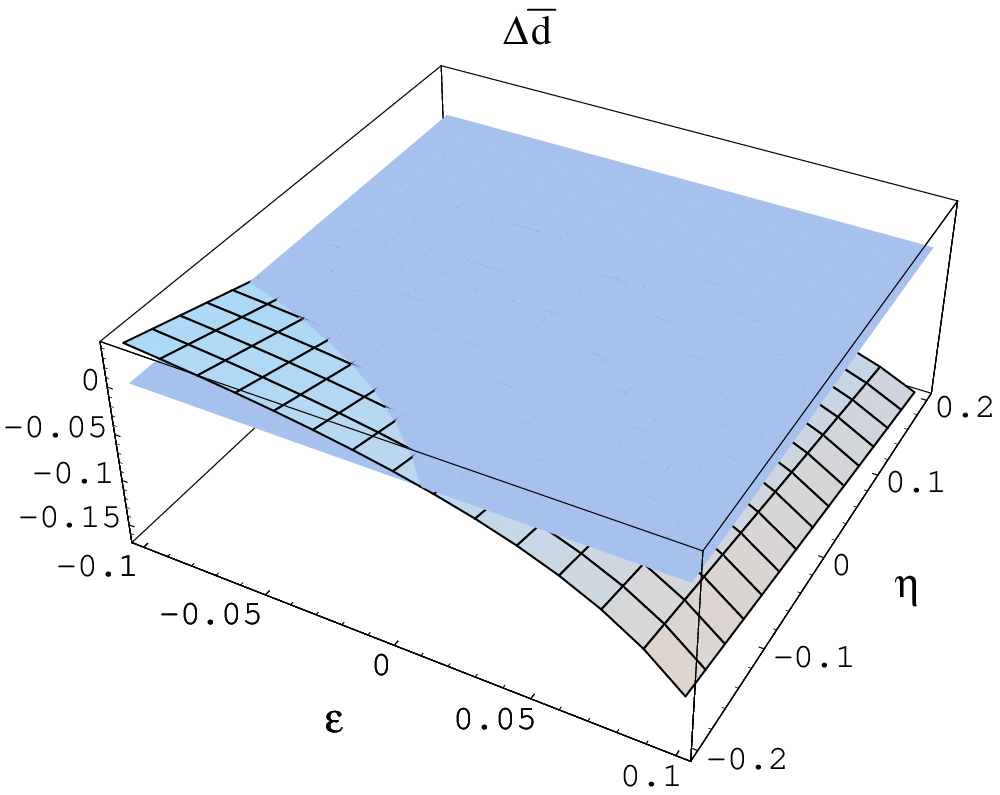}}

\begin{picture}(400,20)
\put(100,10){(a )} \put(300,10){(b)}
\end{picture}

\noindent \hspace{-0.5cm} \scalebox{0.7}{
\includegraphics{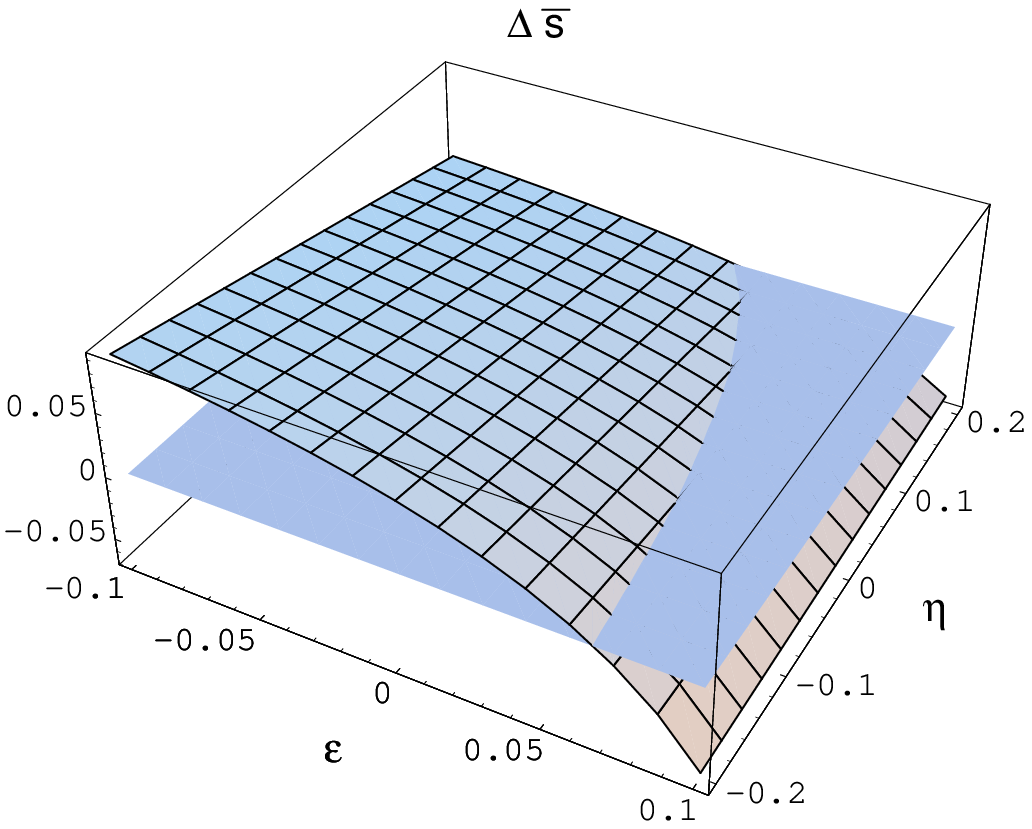}}

\begin{picture}(400,20)
\put(100,10){(c)}
\end{picture}

\caption{\label{fig1}The anti-quark polarizations for $\bar{u}$
(a), $\bar{d}$ (b) and $\bar{s}$ (c) as a functions of parameters
$\epsilon$ and $\eta$ for $\Delta \Sigma=0.3$. For comparison the
plain $\delta \bar{q}=0$ is also shown. }

\end{figure}

\begin{figure}

\noindent \scalebox{0.8}{ \includegraphics{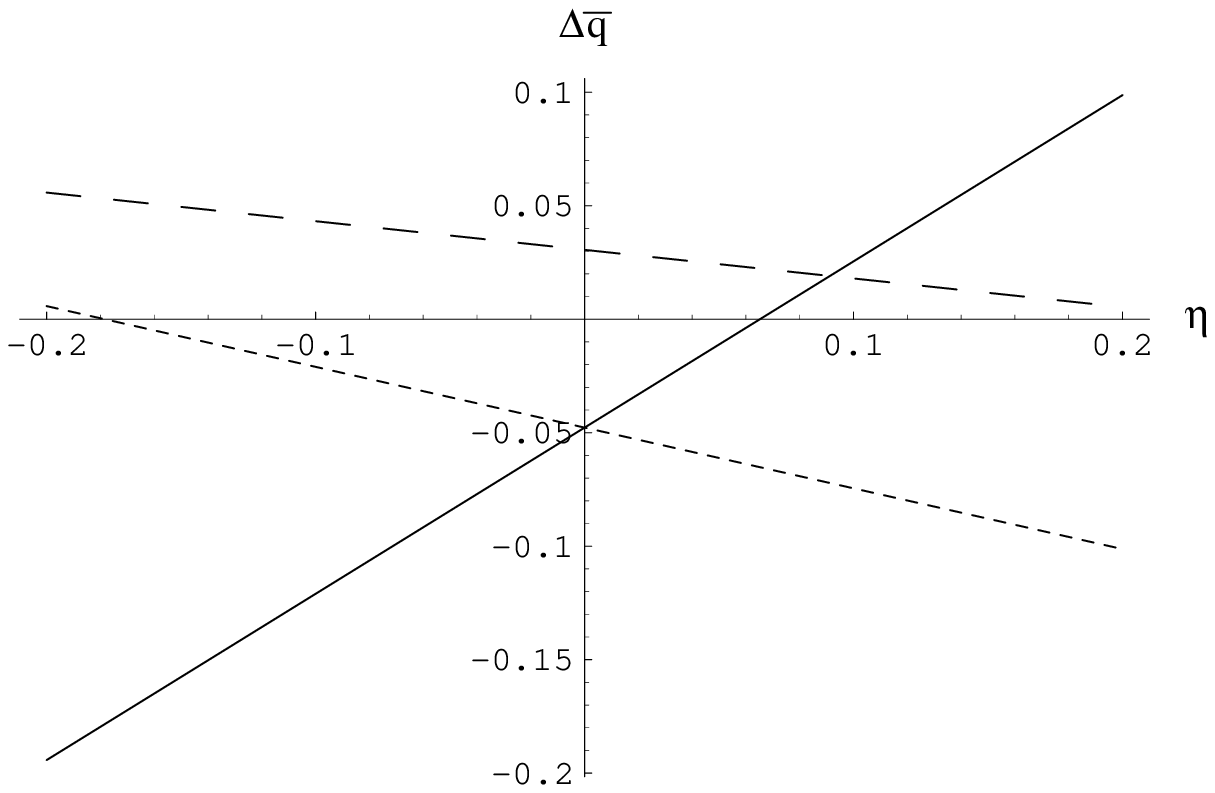}}

\caption{\label{fig2}The anti-quark polarizations for $\bar{u}$
(solid), $\bar{d}$ (short dashed) and for $\bar{s}$ (dashed)
versus $\eta$ for $\epsilon=0$ and $\Delta \Sigma=0.3$.}
\end{figure}

\begin{figure}
\noindent \scalebox{0.8} { \includegraphics{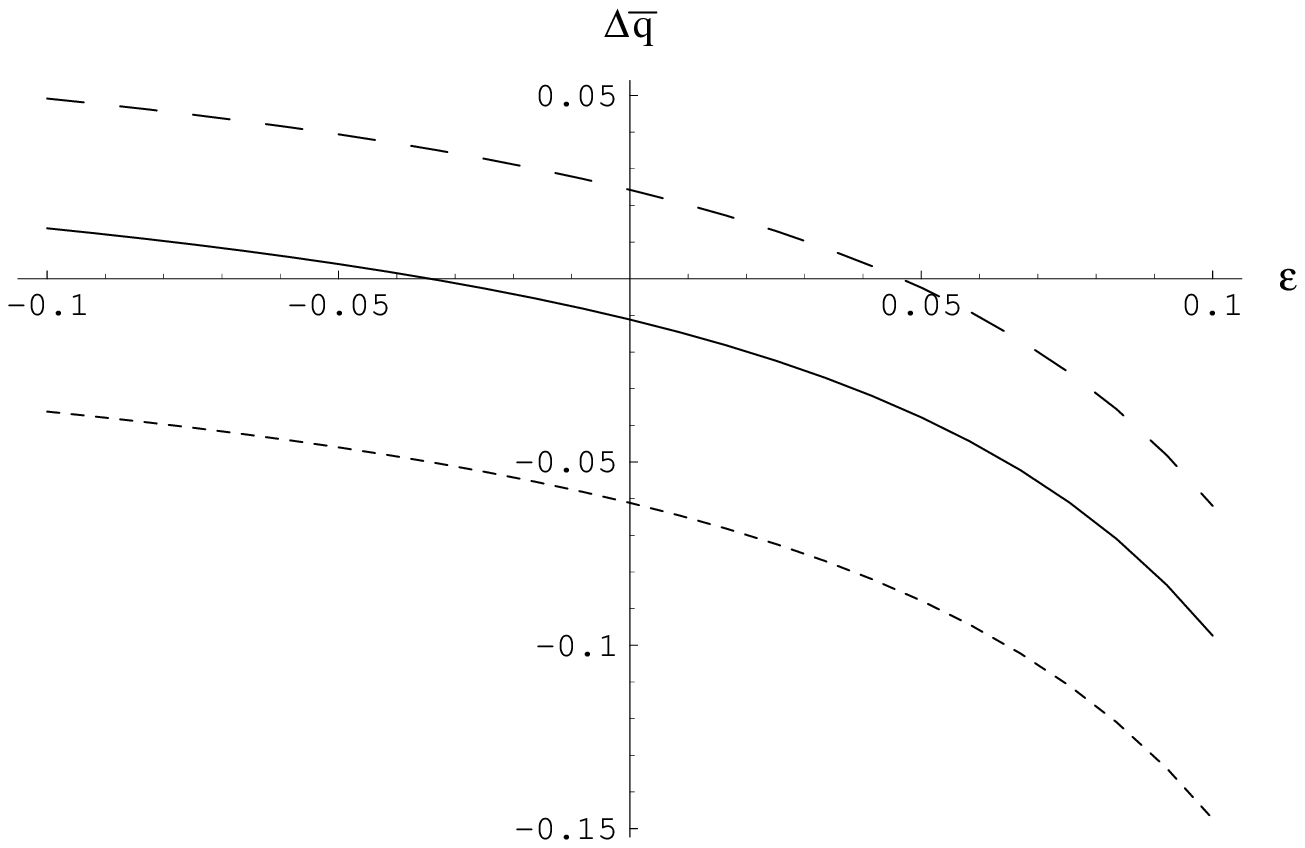}}

\caption{\label{fig3}The anti-quark polarizations for $\bar{u}$
(solid), $\bar{d}$ (short dashed) and for $\bar{s}$ (dashed)
versus $\epsilon$ for $\eta=0.05$ and $\Delta \Sigma=0.3$.}
\end{figure}

Our antiquark polarizations $\Delta\bar{q}$ depend on three
variables: $\epsilon, \, \eta$ and $\Delta \Sigma$. In
FIG.~\ref{fig1}-FIG.~\ref{fig3} we show this dependence with fixed
$\Delta \Sigma=0.3$. In FIG.~\ref{fig1} we present $\Delta
\bar{q}$ for three flavors as a functions of both variables
$\epsilon$ (in a range $-0.1 \leq \epsilon \leq 0.1$) and $\eta$
($-0.2 \leq \eta \leq 0.2$). For comparison the plain
corresponding to $\Delta \bar{q}=0$ is also shown.

The results for $\Delta \bar{q}(\eta)$ for $\epsilon=0$ are given
in FIG.~\ref{fig2}. The changes of $\Delta \Sigma$ shift the whole
diagram parallel to $\Delta \bar{q}$ axis. In the region $-0.2\leq
\eta \leq 0.2$ $\Delta \bar{d}$ and $\Delta \bar{s}$ change slowly
with $\eta$ and dependence of $\Delta \bar{u}$ on this variable is
stronger. As is seen from FIG.~\ref{fig2}  $\Delta \bar{u}$ and
$\Delta \bar{s}$ are not very much different from zero around
$\eta=0.05$ and $\Delta \bar{d}$ is  negative in order to satisfy
the sum rule from Eq.(28).

The dependence of $\Delta \bar q$ on $\epsilon$ is shown in
FIG.~\ref{fig3} for $\Delta \Sigma=0.3$ and $\eta=0.05$ (for
$\eta=0$ $\Delta \bar d$ is identical with $\Delta \bar u$). In
the interesting region $-0.1\leq \epsilon \leq 0.1$ curves for
$\Delta \bar{q}$ are nearly parallel and for higher values the
dependence on $\epsilon$ becomes stronger.

 If we take $\epsilon=0$, $\eta=\Delta_{H}$ and
 $\Delta \Sigma=0.3 \pm 0.1$ we can determine polarizations of
sea antiquarks:
\begin{eqnarray}
\Delta \bar{u} &=& -0.01 \pm 0.05 \, , \nonumber \\
\Delta \bar{d} &=& -0.06 \pm 0.03 \, , \\
\Delta \bar{s}&=& \hspace*{0.33cm}0.02 \pm 0.02 \, .\nonumber
\end{eqnarray}
The values of quoted errors are dominated by the contribution from
the error of $ \eta$. As one can see from Eq.(31) the values of
$\Delta \bar{u}$ and $\Delta \bar{s}$ are consistent with zero
within one standard deviation.  The errors both in $\Delta \Sigma$
and $\eta$ were taken to take into account spread of different
results for these quantities so the errors in  $\Delta \bar{q}$
could be treated for in the same way. From Eqs.(30) and (31) we
can calculate the polarization of $s$ sea quarks. One gets $\Delta
s_{sea} = -0.12 \pm 0.02$, hence $\Delta s_{sea} \neq \Delta
\bar{s}$ in this case. Getting the precise values of
$\Delta\bar{s}$ and $\Delta s_{sea}$ would be interesting for
recent discussion of strangeness in the nucleon \cite{ellis}.

For $\epsilon=0$ and $\eta=0.05 \pm 0.07$ we get for quark magnetic moments
\begin{eqnarray}
\mu_u &=&  \hspace*{0.33cm}2.39 \pm 0.29 \,n.m. , \nonumber \\
\mu_d &=&-1.19 \pm 0.14 \,n.m. , \\
\mu_s &=& -0.79 \pm 0.09 \, n.m.,\nonumber
\end{eqnarray}
whereas for magnetic densities one obtains
\begin{eqnarray}
\delta u &=&  \hspace*{0.33cm}0.86 \pm 0.10 \, , \nonumber \\
\delta d &=& -0.31 \pm 0.04 \, , \\
\delta s &=& -0.15 \pm 0.02 \, .\nonumber
\end{eqnarray}

For such value of the parameter $\epsilon$   we have for $s$
flavor contribution (which do not depend on $\eta$) to nucleon
moments: $\mu_s \delta s = 0.116 \pm 0.004$ n.m. .

In the given formulas the possible errors connected with the value
of $\epsilon$ are not included.  Comparing quark densities
calculated from magnetic moments with those given in Eq.(30)
coming from DIS we see that with our choice of $\epsilon$ and
$\eta$ the main difference is for $d$ antiquarks. The antiquark
polarizations can not be identical for all flavors because of the
sum rule given in Eq.(28), but the different choice of $\epsilon$
and $\eta$ can give other antiquark polarizations than in Eq.(31).

For the another considered by us as rather exotic choice of
parameter $\epsilon=\epsilon_0$, i.e. such that gives $\delta
s=0$, we get the following values of an antiquark polarizations:
\begin{eqnarray}
\Delta \bar{u} &=& -0.08 \pm 0.06 \, , \nonumber \\
\Delta \bar{d} &=& -0.13 \pm 0.02 \, , \\
\Delta \bar{s}&=& -0.05 \pm 0.02 \, .\nonumber
\end{eqnarray}

All numbers are smaller by 0.07 than the ones given in Eq.(31)
because we have $\Delta \bar{u} (\epsilon=\epsilon_0)-\Delta
\bar{u} (\epsilon=0)= g[f(\epsilon=0)-f(\epsilon=\epsilon_0)]/12
r$ (provided we do not change $\eta$ and $\Delta \Sigma$). In the
case when $\delta s=0$
  we get $\Delta \bar{s}=\Delta s_{sea}=-0.05$.

In \cite{karl,casehg} in the fit for magnetic moments $\delta
q=\Delta q$ is used. The $\chi^{2}/d.o.f.$ is small because one
uses artificial errors ($\pm 0.1$ n.m.) instead of experimental
ones. For such fit one is not able to get the values of antiquark
polarizations.

If we include orbital moments in our analysis $\delta q$ is given
by Eq.(23) (see e.g. \cite{song})

Our sum rule (Eq.(28)) is also changed
\begin{equation}
 \Delta \bar{u}- \Delta \bar{d}-\frac{2r}{r+1} ( \Delta \bar{u}- \Delta \bar{s})=\frac{g_{A}-ra_{8}}{2(r+1)}+\Delta \it{L} \, ,
\end{equation}
where
 \begin{eqnarray}
\Delta \it{L}&=&(<\hat{L}_{z}^u>-<\hat{L}_{z}^{\bar u}>)- (<\hat{L}_{z}^d>-<\hat{L}_{z}^{\bar d}>) \nonumber \\
&&-\frac{2r}{r+1}[(<\hat{L}_{z}^u>-<\hat{L}_{z}^{\bar u}>)-(<\hat{L}_{z}^s>-<\hat{L}_{z}^{\bar s}>)] \, .
 \end{eqnarray}

Hence, it  is not possible to determine antiquark polarizations in
nucleon without any knowledge about angular momenta of quarks. In
the first part of this paper we have assumed
$<\hat{L}_{z}^q>=<\hat{L}_{z}^{\bar q}>$, now we shall try to get
the results with a specific model for these angular momenta. There
is a possibility to improve our fit to magnetic moment by taking
into account another phenomenological contribution similar to
collective orbital momenta of Casu and Seghal \cite{casehg}. It
could be that in such model quarks and antiquarks rotate with
orbital momentum $L$. Our formulas for magnetic moment get an
additional contributions
 \begin{equation}
 \mu (B)=\ldots+\frac{e_B}{2m_B}\it{L} \, .
 \end{equation}

The fit is excellent in this case an one gets $\chi^{2}/d.o.f. = 0.06$ and the parameters does not change very much in comparison with the
previous fit
\begin{eqnarray}
c_{0} &= & 0.042 \pm 0.007\, n.m. \, , \nonumber \\
c_{3} &=& 1.037 \pm 0.007\, n.m. \, , \nonumber \\
c_{8} & =& 0.179 \pm 0.009\, n.m. \, , \\
r & =& 1.465 \pm 0.047\,  , \nonumber \\
V &=& 0.24 \pm 0.02\, n.m. \; , \nonumber \\
L &=& 0.08 \pm 0.05 \; . \nonumber
\end{eqnarray}

The resulting values for magnetic moments of octet baryons are
presented in Table \ref{tab1} (model B).

One can repeat the calculations done before with new parameters.
 The value of right-hand side in Eq.(28) of our basic sum rule changes from
0.09 to $0.08 \pm 0.01$. The values of antiquark polarizations
(for $\epsilon=0$, $\eta=0.05 \pm 0.07$ and $\Delta \Sigma =0.3
\pm 0.1$) does not change significantly in comparison to the ones
presented in Eq.(31):
\begin{eqnarray}
\Delta \bar{u}&=& \hspace*{0.33cm}0.01 \pm 0.05 \, , \nonumber \\
\Delta \bar{d}&=& -0.04 \pm 0.03 \, , \\
\Delta \bar{s}&=& \hspace*{0.33cm}0.03 \pm 0.02 \, .\nonumber
\end{eqnarray}

Summarizing, we have modified generalized Sehgal equations for
magnetic moments of baryons and we get the very good fit using
experimental errors. With 4 free parameters in this fit we are not
able to determine 6 quantities, namely 3 magnetic moments of
quarks and 3 quark densities. We get sum rules for the ratios.
Using information on deep inelastic scattering of polarized
particles and $\beta$-decays and connecting quark densities from
magnetic moments with those from spin asymmetries we can express
anti-quark densities as function of two parameters $\epsilon$ and
$\eta$. We give antiquark polarizations calculated with the
assumption that $\mu_u/\mu_d=-2$,  i.e. $\epsilon=0$ and
$\eta=0.05$ (value given in results of Hermes experiment). Taking
into account errors the results are not very conclusive but
because of very weak dependence on the parameters it seems that
$\bar{u}$ and $\bar{s}$ are close to zero  and $\bar{d}$ is small
and negative. To really calculate the antiquark polarizations
additional precise information on quark magnetic moments and quark
densities is needed. By taking very specific corrections connected
with orbital angular momentum proportional to the charge of the
baryon we can get nearly perfect description of baryon octet
magnetic moments. These corrections are not big and do not change
conclusions from the first part of the paper.

\end{document}